\documentclass[aps,prl,amssymb,floats,twocolumn]{revtex4}

\usepackage{epsfig}
\usepackage{color}
\pagecolor{white}
\usepackage{latexsym}
\usepackage{hyperref}

\begin{document}


\title{Upper limits on gravitational waves emission in association with the Dec 27 2004 giant 
flare of SGR1806-20 }
\author{
L. Baggio$^1$, M. Bignotto$^2$, M. Bonaldi$^3$, M. Cerdonio$^{\ast 2}$, 
\thanks{The AURIGA collaboration; corresponding author: cerdonio@pd.infn.it}
L. Conti$^2$, M. De Rosa$^4$,  P. Falferi$^3$,  P. Fortini$^5$,  M. Inguscio$^6$, 
N. Liguori$^2$, F. Marin$^6$,
 R. Mezzena$^7$,  A. Mion$^7$, A. Ortolan$^8$, G.A. Prodi$^7$, S. Poggi$^7$,
F. Salemi$^6$, G. Soranzo$^{9}$, L. Taffarello$^{9}$, G. Vedovato$^{9}$, A. Vinante$^7$,
S. Vitale$^7$ and J.P. Zendri$^{9}$ \\ (the AURIGA Collaboration)$^{10}$ \\ 
}

\address{$1$ Institute for Cosmic Ray Research, Univ. of Tokyo,
5-1-5 Kashiwanoha, Kashiwa, Chiba, 277-8582, Japan}

\address{$^2$INFN\ Padova Section and Department of Physics, University of
Padova,  I-35131 Padova,  Italy}

\address{$3$ Istituto di Fotonica e Nanotecnologie CNR-ITC and
INFN Gruppo Collegato di Trento, Padova Section, I-38050 Povo (Trento), Italy}

\address{$4$ INOA I-80078 Pozzuoli (Napoli),  Italy and INFN Firenze Section, I-50121 Firenze, Italy}

\address{$5$ Physics Department, University of Ferrara and INFN Ferrara Section, I-44100 Ferrara, Italy}

\address{$6$ LENS and Physics Department, University of Firenze and INFN Firenze Section, I-50121 Firenze, Italy}

\address{$7$ Physics Department, University of Trento and INFN Gruppo Collegato di Trento, 
Padova Section, I-38050 Povo (Trento), Italy }

\address{$^8$INFN, Laboratori Nazionali di Legnaro,  I-35020 Legnaro (Padova) Italy}
\address{$^9$ INFN Padova Section, I-35100 Padova, Italy}

\address{$^{10}$ http://www.auriga.lnl.infn.it}

\date{$\,^*$Corresponding author: cerdonio@pd.infn.it}

\begin{abstract}
At the time when the giant flare of SGR1806-20 occurred, the AURIGA ``bar" 
gw detector was on the air with a noise performance close to stationary gaussian. 
This allows to set relevant upper limits, at a number of frequencies 
in the vicinities of  900 $Hz$, on the amplitude of the damped gw wave trains,  
which, according to current models,  could have been emitted, due to the 
excitation of normal modes of the star associated with the peak in X-rays 
luminosity.
\end{abstract}

\maketitle
\


On  27 December 2004 the Soft Gamma-ray Repeater SGR1806-20 gave a giant 
flare, which was observed by a number of instruments \cite{Borkowsky}.

The  fluence,  if the emission is assumed isotropic,  at the distance of $d\sim 15 \ kpc$ 
would imply an energy some hundred times larger than any other known giant 
flare \cite{Hurley, Terasawa}. 
Soft gamma-ray repeaters are thought to be magnetars (see \cite{Hurley} and refs. 
therin). 
It has been suggested \cite{Hurley,Schwartz} that the extreme energy event of 27 December 
2004 is due to a catastrophic instability  involving global crustal 
failure and magnetic reconnection \cite{Thomson1}.
Observations by CLUSTER and TC-2, in combination of data from GEOTAIL, 
gave evidence that the steep initial rise contains two exponential phases, of 
e-folding times 4.9 $ms$ and 67 $ms$ respectively, which covered the 24 $ms$ 
before the time of the peak intensity $t_{p}$; 
all the timescales support the 
notion of a sudden reconfiguration of the stars magnetic field, producing large 
fractures in the crust \cite{Schwartz}. In particular these authors remark that the 
intermediate $\approx 5\ ms$ time is naturally explained if the rising time 
is limited by ithe propagation of a triggering fracture of size 
$\approx 5 \ km$, as it would be predicted by the theory of reference \cite{Thomson2}.
 
According to a few somewhat different models, 
as a consequence of crustal cracking \cite{de Freitas} or reconfiguration of the moment of inertia tensor 
\cite{Ioka}, non-radial kHz oscillation 
 modes of the neutron star would be excited, giving emission of gravitational waves (gw), possibly at frequencies
  where the gw bar detector AURIGA \cite{AURIGA} is sensitive (see insert of Fig. 1). 
Both the above quoted models predict gw emission, starting very close to $t_p$,   
which involves $kHz$ non radial modes of oscillation of a neutron 
star with few hundred $ms$ damping time. The expected waveforms can be 
approximately parametrized
as $h(t)=h_0\exp(-t/\tau_s)\sin(2 \pi f_s t)$, where 
$h_0$ is the maximum gw amplitude, $f_s$ and $\tau_s$ 
are the frequencies and damping times of normal modes; 
the polarization of the wave is not known. 
The frequencies of the various modes are still under study and 
depend on a variety  of factors as EOS, temperature, density, age, 
rotational state of the star, etc. \cite{miniutti} so that we are unable to 
anticipate with any confidence what specific set of gw emission frequencies 
could be the one expected for a magnetar ready to undergo a supergiant flare. 
Still the lowest lying modes, $g-$,  $f-$ and, marginally, $p-modes$, 
could well be in the frequency range $500\div 1500 \ Hz$, depending on the status of the star.

Within a factor 10 in gw amplitude, AURIGA is sensitive to gws from $\sim$ 800 
to 1050 $Hz$. Here we limit the analysis to the most sensitive part of the 
band, namely between $850$ and $950 \ Hz$ (see the insert in Fig. 1), 
where the detector sensitivity varies no more than a factor 4 in amplitude. 
Since the  last upgrading of the suspensions on Dec $2^{nd}$ 2004 the detector 
is well behaved in the sense that performs stationary gaussian, after 
epochs of environmental disturbances are vetoed  by means of auxiliary channels 
(i.e. signals at frequencies where the detector is gw insensitive). 
During nights and week-ends the 
vetoed epochs becomes less frequent and shorter, so that the detector 
achieve close to 90\% stationary  gaussian operation; 
in particular on time spans of minutes we can use the  data, without even 
applying vetoes. 
This is the case for the time span of about $\pm 100 s$ around
 the epoch of the 27 December 2004 giant flare of SGR1806-20, which we use 
in this analysis. 
\begin{figure} [hbt]
\begin{center}
\epsfxsize=1.0\linewidth\epsffile{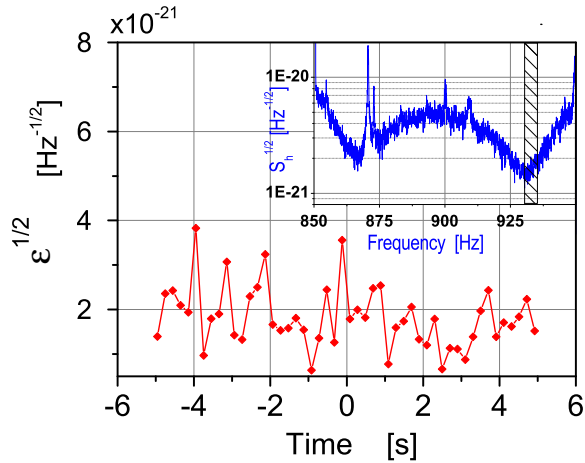}
\end{center}
\caption{\label{f1}  Plot of ${\cal E}^{1/2}$ 
in the frequency band $930 \div 935 \ Hz$ as a function of time; the origin in the x-axis corresponds 
to the arrival time of the flare X of SGR1806-20 at the AURIGA site. 
The insert shows the AURIGA one-sided noise
spectral density, as expressed in equivalent gw amplitude $h_r(t)$ at input. The vertical dashed
area shows the position of the frequency bin $930 \ Hz$ (see text).}
\end{figure}
We show in the following that the noise  
is driven by a zero mean stochastic gaussian 
process with a stationary correlation function. 
For what concern the directional sensitivity, 
the orientation of AURIGA in respect to the direction of SGR1806-20 
was such that the antenna pattern, averaged over polarizations, 
gave maximal sensitivity at the time of the giant flare.
Then we have a unique opportunity to search in our data for gravitational waves emitted at the peak time of the
giant flare. We take the peak time $t_p$ to be 21:30:26.68 UT 
of 27 December 2004 after taking into account the time difference
between the arrival time at INTEGRAL \cite{int} and at AURIGA site of 
$133.427\ ms$.\cite{Lichti} This time corresponds also to the
peak position of the CLUSTER data which show, 
after the last exponential rise, the evident start of a phase in which damping 
occurs until the signal gets below $1/10$ of the peak value, $\sim 300 \ ms$ 
after the peak.\cite{Schwartz} Following the models quoted 
above, in both cases we can assume the peak time $t_p$ as the start of the gw 
excitation and $\tau_s$ =  100 $ms$, that is 1/3 of  300 $ms$, as the
corresponding damping time. 
In order to extract the signal power first we reconstruct the gw amplitude $h_r(t)$ at input
through the detector transfer function. Then we slice the gw sensitive frequency band
of AURIGA in contiguous and non-overlapping sub-bands $f_j$ of constant width $\Delta f$, and 
centered in $f_j^c=f_j+\Delta f/2$, by means of 
digital top-hat filters in the frequency domain $T_j(f)= \vartheta(|f|-f_j) - 
\vartheta(|f|-f_j-\Delta f)$. 
Within each sub-band, we compute the the equivalent input signal power over a time span $\Delta t$
\begin{equation}
{\cal E}_j \equiv \int_{\Delta t}\ T_j\ast h_r^2(t+k\Delta t)\ dt \ ,
\end{equation}
where $\ast$ stands for time convolution.  The ${\cal E}_j(t)$ is sampled every 
$\Delta t$ to construct the time series ${\cal E}_j(k)$ with $k$ integer.
We decided {\it a priori} a fixed partition of the time frequency plane: 
$\Delta f = 1/(2 \tau_s) = 5 \ Hz$ and $\Delta t = 201.5 \ ms \simeq 2 \tau_s$. 
For each sub-band $f_{j}$, 
we analyzed the resulting time series of ${\cal E}_j(k)$ over a time span of $\pm 100 \ s$ around 
the peak time $t_p$ to check the ``off source" noise statistics. The
${\cal E}_j(k)$ sample including the peak time $t_p$ is then compared to 
the measured noise statistics, looking for any evidence of excess power. 
To be more precise, the {\it a priori} choice of our sampling time made $t_p$ to
fall $120\ ms $ after the beginning of the integration time $\Delta t$ of the 
``on source" sample. Fig. 1 shows how ${\cal E}_j$ fluctuates on the time spans
of $\pm 5 \ s$ around the time of the flare $t_p$ for the sub-band $f_j=930\ Hz$.
A gw emission at frequency $f_s$ would give an excess power in the band $\Delta f$ centered 
at the $f^c_j$ such that $|f_s-f^c_j|< \Delta f/2$. The released energy would 
be maximum in the ``on source" sample. The excess signal power in each sub-band 
$\Delta f$ can be easily calculated from the expected waveform and reads
\begin{eqnarray}
{\cal E}_{s} &\simeq & (h^2_0  \tau_s/4) \ \biggl\{\biggl[\frac{1}{2}\, +\, \frac{1}{\pi}\tan^{-1}\biggl(\frac{2
x}{\delta^2-x^2}\biggr)\biggr] \nonumber \\
&+& O\biggr(\frac{1}{f_s\tau_s}\biggl)^2\biggr\} \ ,
\end{eqnarray}
where $x\equiv (2 \pi \tau_s)^{-1}/\Delta f$ and $\delta\equiv |f_s-f_j^c|/\Delta f$
are the ratios between the signal bandwidth ($\equiv 1/(2 \pi \tau_s)$)
and the detuning of the signal frequency and the bandwidth $\Delta f$, respectively.
With our choice of parameters for the analysis, the excess signal power is approximately 
(within a few \% error)
\begin{equation}
{\cal E}_{s} \approx \frac{h^2_0\tau_s}{6} \biggl[1-\biggl(\frac{f_s-f_j^c} {\Delta f_{eff}}\biggr)^2 \biggr]\ ,
\end{equation}
where $\Delta f_{eff}= 4 \ Hz$. 
To check the statistics of the ``off source" samples,  we histogram each time series 
${\cal E}_j(k)$ and compare them with the predicted probability density functions 
assuming gaussian noise, by fitting for the variance separately in each sub-band.
The fitting probability density function is a $\chi^2$ distribution
with $\alpha$ effective degrees of freedom
$p({\cal E};\sigma^2)= 2^{-\alpha/2}\, ({\cal E}/\sigma^2)^{\alpha/2-1}
\exp(-{\cal E}/2 \sigma^2)/\Gamma(\alpha/2)/\sigma^2$,
where $\sigma^2$ is the variance of the underlying gaussian stochastic process.
We show in Fig. 2 the close agreement with prediction of the data for the 
frequency bin $f_j = 930\ Hz$, over the extended time span of $\pm 100 \ s$.
\begin{figure} [hbt]
\begin{center}
\epsfxsize=1.0\linewidth\epsffile{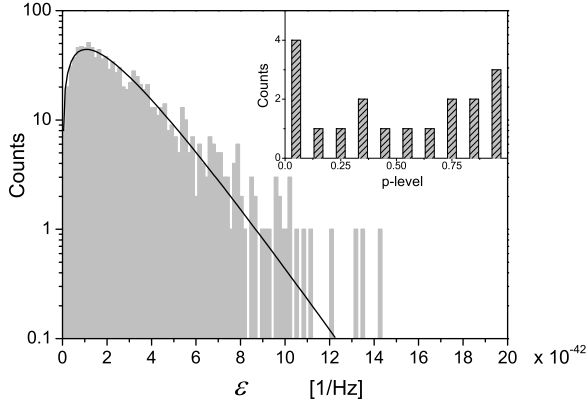}
\end{center}
\caption{\label{f1}  Histogram of ${\cal E}$ in the frequency band
$930 \div 935 \ Hz$; solid line represents the best fit curve of  
a $\chi^2$ distribution with $\alpha=3.6$ effective degrees of freedom. The insert
shows the p-level distribution of the same fit for the histograms of the 18 frequency bins.}
\end{figure}
The results for all other frequency bins are similar. The
goodness of the fit has been checked by a $\chi^2$ test, and 
the resulting p-values for all the sub-bands are
consistent with a uniform distribution in the unit interval, 
as expected (see the insert of Fig. 2 and Tab.1). 
In Table 1 we report the parameter $\sigma^2$ 
of the fit of $p({\cal E})$ to the experimental data.   
We find that the dependence on the effective degrees of freedom $\alpha$ of 
the $p$-levels is very weak and, within the statistical errors, 
we can fix $\alpha=3.6$ for all the frequency bins. The $p$-level distribution is 
 uniform in the unit interval (see the insert of Fig. 2). 
The stationary behavior, at least for timescales of few minutes, 
is shown by the constancy in time of the parameters needed to fit the noise model. 
\begin{table}[!ht]
\caption{List of fit parameter $\sigma^2$ of the
histogrammed  ${\cal E}$ data samples $\pm 100\ s$ around $t_p$ tabulated as 
increasing sub-band frequencies; the data for the $870 \ Hz$ sub-band have been 
discarded {\it a priori} as this band is contaminated by environmental noise.
The ``on source" value of ${\cal E}$ including the trigger time $t_p$ is also reported 
as well as the computed upper limit with confidence $\ge 95\% $.}

\begin{center}
{\small
\begin{tabular}{cccccc}
\tableline
\noalign{\smallskip}
$f_j \ $ &  $\ \ \ \sigma_j^2 \times 10^{42}\ \ \ $  &  $ \ {\cal E}_{t_p} \times 10^{42}\ $ &  $ \ {\cal E}_{95} \times 10^{40}\ $\\
$[Hz]$ &  $[Hz^{-1}]$ &   $[Hz^{-1}]$  &  $[Hz^{-1}]$\\
\noalign{\smallskip}
\tableline
\noalign{\smallskip}
855 &  $4.78$  & $4.90$ & $0.86$ \\
860 &  $1.89$  & $4.15$ & $0.34$ \\
865 &  $1.96$  & $9.83$ & $0.35$ \\
875 &  $2.94$  & $2.93$ & $0.53$ \\
880 &  $4.30$  & $10.3$ & $0.77$ \\
885 &  $5.11$  & $4.52$ & $0.92$ \\
890 &  $6.15$  & $6.51$ & $1.11$ \\
895 &  $5.89$  & $8.57$ & $1.06$ \\
900 &  $6.93$  & $6.60$ & $1.25$ \\
905 &  $6.18$  & $6.78$ & $1.11$ \\
910 &  $3.69$  & $19.7$ & $0.66$ \\
915 &  $2.60$  & $7.06$ & $0.47$ \\
920 &  $1.61$  & $6.57$ & $0.29$ \\
925 &  $0.87$  & $3.19$ & $0.16$ \\
930 &  $0.71$  & $5.25$ & $0.13$ \\
935 &  $1.24$  & $2.57$ & $0.22$ \\
940 &  $3.56$  & $19.3$ & $0.64$ \\
945 &  $11.2$  & $30.2$ & $2.01$ \\

\noalign{\smallskip}
\tableline
\end{tabular}
}
\end{center}
\end{table}
We take advantage of the classical theory of hypothesis testing to
establish if the samples ${\cal E}_{t_p}$ corresponding to the arrival time
$t_p$ are affected by the presence of a gw signal.
To test the null hypothesis ${\cal H}_0$, i.e. that the sample is drawn from the
estimated noise probability distribution in absence of signals, we set a
threshold ${\cal E}_{cr}$ corresponding to a confidence level (C.L.)
$p({\cal E}<{\cal E}_{cr})\geq 1-p_{cr}$. The threshold for $1-p_{cr}=95$ \% C.L.
corresponds to ${\cal E}_{cr}=8.8\times \sigma_j^2$.
Thus one sees from Table 1 that no excess of gw power is found at $t_p$ and 
therefore we have to set up upper limits.
We set conservative confidence intervals for ${\cal E} _s$ using a confidence belt 
construction \cite{hagi} which ensures non-uniform coverage greater or equal to 90\%. 
The confidence belt construction proceeds as follows.
Assume that the signal magnitude is ${\cal E} _s$. The measured ${\cal E}$ in each sub-band 
(Eq. 1) obeys a non-central $\chi^2$ distribution with central parameter equal to 
${\cal E} _s/\sigma^2$ (here we drop for simplicity the index of the sub-band). 
Its corresponding probability density function can be written as:
\begin{eqnarray}
p({\cal E} ;{\cal E} _s ,\sigma ) &=& \frac{1}{2 \sigma^2}\exp \left( { - \frac{{{\cal E}  + {\cal E} _s }}{{2\sigma^2 }}} \right)\left( {\frac{{\cal E} }{{{\cal E} _s }}} \right)^{(\alpha  - 2)/4} \times \nonumber \\ 
&\times & I_{{\alpha/2  - 1} } \left( {\sqrt {{\cal E} {\cal E} _s /\sigma^2 } } \right) \ , 
\end{eqnarray}
where $I_k(x)$ are the modified Bessel functions of the first kind of order $k$.
The $q$-quantile of this distribution, ${\cal E} _q({\cal E} _s ,\sigma )$, is implicitly defined by
$q = \int_0^{{\cal E} _q } {p({\cal E} ;{\cal E} _s ,\sigma )d{\cal E} }$.
For each value of the unknown ${\cal E} _s$ we define the 95\% confidence belt
boundaries ${\cal E}_{hi}$ and ${\cal E}_{low}$ as
\begin{equation}
\begin{array}{l}
 {\cal E} _{hi } ({\cal E} _s ,\sigma ) = \left\{ {\begin{array}{*{20}c}
   0 & {{\rm if}\quad {\cal E} _s  < {\cal E} _s^{cr} (\sigma )}  \\
   {{\cal E} _{5\% } ({\cal E} _s ,\sigma )} & {{\rm otherwise}}  \\
\end{array}} \right. \\
 {\cal E} _{low } ({\cal E} _s ,\sigma ) = {\cal E} _{95\% } ({\cal E} _s ,\sigma ) \\
 \end{array}
\end{equation}
where ${\cal E} _s^{cr}$ is implicitely defined by
${\cal E} _{5\% } ({\cal E} _s^{cr} ,\sigma ) = {\cal E} _{95\% } (0,\sigma )$.
This confidence belt defines a set of confidence intervals on ${\cal E} _s$,
whose frequentist coverage is -- by construction -- 90\% for
${\cal E} _s>{\cal E} _s^{cr}$, and 95\% for ${\cal E} _s \le {\cal E} _s^{cr}$.
In other words, for every value of ${\cal E} _{t_p}$ from each sub-band,
if ${\cal E} _{t_p}<{\cal E} _{95\% } (0,\sigma _j )$ we set an upper limit 
equal to ${\cal E} _s^{cr}$, otherwise our procedure gives a two-sided 
confidence interval. In all sub-bands we obtain
upper limits, which can be written as ${\cal E} _s^{cr} \simeq 18 \times \sigma_j^2$.
These limits ranges from ${\cal E}^{1/2} = 3.5\times 10^{-21}Hz^{-1/2}$
to ${\cal E}^{1/2} = 1.4\times 10^{-21}Hz^{-1/2}$, according to AURIGA sensitivity.

The initial amplitude of the neutron star normal modes $h_0$ is related to 
${\cal E}$ by Eq. (3) that gives, for the best upper limit, 
$h_0\leq 2.7 \times 10^{-20}$.
We discuss now the upper limit in terms of the 
total  gw energy $\epsilon_{gw}=E_{gw}/M_\odot c^2$ emitted by the 
normal modes excitation during the peak of the giant flare of SGR1806-20. 
The well known formula of the quadrupolar radiation, for the 
expected gw signal \cite{de Freitas}, can be written as $h_0=(\epsilon_{gw}\ c 
R_S/(4 \pi^2 \tau_s)^{1/2}/(f_s d)$, where $R_S$ is the
Swartzchild radius of one solar mass black hole.  
Thus the resultant upper limit on $\epsilon_{gw}$ reads  
\begin{eqnarray}
\epsilon_{gw}  &\leq & 3 \times10^{-6}\biggl(\frac{{\cal E}}{1.3 \times 10^{-41} Hz^{-1}}\biggr)\times \nonumber\\ 
&\times &
\biggl(\frac{15\ kpc}{d} \biggr)  
\biggl( \frac{930 \ Hz }{f_s} \biggr)^2  \biggl( \frac{\tau_s}{0.1 s}\biggr)\ .
\end{eqnarray}
We should notice that a gw bar detector has a polarization dependent 
sensitivity; hence, for an unpolarized or linearly polarized gw, 
the result in Eq. (6) should be multiplied by a 
factor $2$ or $\cos^2(2 \psi)$ respectively, where $\psi$ is the angle 
between the bar axis and the polarization of the wave.
We conclude that, if the star ever emitted gws from excitation of its normal modes 
at any of the frequencies studied here, in the time span $\Delta t$ containing the 
flare time $t_p$, the gw amplitudes and energetics are limited as above.
\begin{figure} [hbt]
 \begin{center}
 \epsfxsize=1.03\linewidth\epsffile{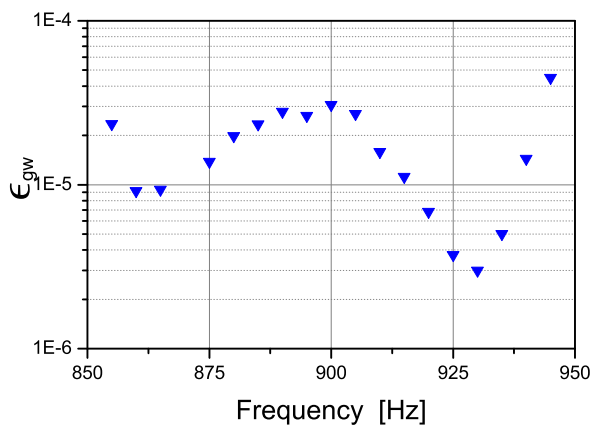}
 \end{center}
 \caption{\label{f2} Upper limts on the gw energy released at the source around the flare
peak time $t_p$ expressed as a fraction of $M_\odot c^2$.
}
 \end{figure}
If the giant flare of SGR1806-20 on 27 December 2004 is indeed 
some 100 times more energetic (however see ref. \cite{Yamazaki}) and if the gw luminosity  
scales with the em luminosity, then,  
 for the frequencies considered,  our upper limits come close  
to the predictions of the models of refs. \cite{de Freitas, Ioka} which give 
an energetics of the order of $\epsilon_{gw} \approx 5 \times 10^{-6}$.
The method used here is of course sub-optimal and the upper limits are somewhat 
weaker than the ``optimal" matched filter. 
In any case this work shows that, 
as there is the specific peak time $t_p$ to be used as external trigger, 
it is worth to make searches 
even with a single detector if its noise is well behaved. 
An extension of such  
searches involving the gw detectors on the air in a coincidence search, 
would also allow to use the information of the gw 
travel delays between the detectors to select against spuria, and would give the 
most exhaustive and efficient search, in terms of frequency coverage and 
confidence in improving the limits, if not to get a candidate detection.
\begin{acknowledgments}
We are gratefull to Roberto Turolla for a critical reading of the manuscript.  
\end{acknowledgments}


\begin{thebibliography}{reference}
\bibitem{Lichti}
We owe to the courtesy of G. Lichti (INTEGRAL collaboration) the calculation of the arrival time of the flare X
at the AURIGA site.
\end{thebibliography}
\end{document}